\documentclass[12pt]{article}

\usepackage{amsbsy}
\usepackage{amsmath}
\usepackage{amssymb}
\usepackage{amsthm}
\usepackage{authblk}
\usepackage{color}
\usepackage{geometry}
\usepackage{graphicx}
\usepackage{caption}
\usepackage{subcaption}
\usepackage{natbib}
\usepackage[hidelinks]{hyperref}
\usepackage{setspace}
\usepackage{mathrsfs}
\usepackage{physics}
\usepackage{float}
\usepackage{comment}
\theoremstyle{plain}
\newtheorem{thm}{Theorem}

\newtheorem{cor}{Corollary}

\def\bp{\mathbf{p}}
\def\bx{\mathbf{x}}
\def\bz{\mathbf{z}}
\def\bI{\mathbf{I}}
\def\bQ{\mathbf{Q}}

\def\bmu{\boldsymbol{\mu}}
\def\bSigma{\boldsymbol{\Sigma}}
\def\bOmega{\boldsymbol{\Omega}}

\begin{document}

\title{Reference-Invariant Inverse Covariance Estimation with Application to Microbial Network Recovery}
\author{Chuan Tian, Duo Jiang, Yuan Jiang\thanks{Yuan Jiang is the corresponding author. This research is supported in part by National Institutes of Health grant R01 GM126549.}}
\affil{Department of Statistics\\ Oregon State University}
\date{\today}
\maketitle

\begin{abstract}
The interactions between microbial taxa in microbiome data has been under great research interest in the science community. In particular, several methods such as SPIEC-EASI, gCoda, and CD-trace have been proposed to model the conditional dependency between microbial taxa, in order to eliminate the detection of spurious correlations. However, all those methods are built upon the central log-ratio (\textsc{clr}) transformation, which results in a degenerate covariance matrix and thus an undefined inverse covariance matrix as the estimation of the underlying network. \cite{jiang2021microbial} and \cite{tian2022compositional} proposed bias-corrected graphical lasso and compositional graphical lasso based on the additive log-ratio (\textsc{alr}) transformation, which first selects a reference taxon and then computes the log ratios of the abundances of all the other taxa with respect to that of the reference. One concern of the \textsc{alr} transformation would be the invariance of the estimated network with respect to the choice of reference. In this paper, we first establish the reference-invariance property of a subnetwork of interest based on the \textsc{alr} transformed data. Then, we propose a reference-invariant version of the compositional graphical lasso by modifying the penalty in its objective function, penalizing only the invariant subnetwork. We validate the reference-invariance property of the proposed method under a variety of simulation scenarios as well as through the application to an oceanic microbiome data set.
\end{abstract}

\section{Introduction}
\label{Sec-3.1}

Microbiome, which is the collection of micro-organisms in an ecological system, is of great research interest in the science community and has been shown to play an important role in influencing its host or living environment. For instance, it is found that intervening the gut microbiota of African turquoise killifish could result in delay of their aging process \citep{smith2017regulation}. The advancement of the high-throughput sequencing technologies such as 16S rRNA profiling that replicates a specific sequence of marker genes which is counted and serves as the proxy of the abundance of the Operational Taxonomic Units (OTU's, the surrogate of bacteria species) in a sample, has enabled researchers to analyze the microbial compositions in uncultivated samples.

One common goal in microbiome data analysis is to understand how microbes interact with each other. However, the nature of microbiome data determined by the technicalities of the sequencing procedures has imposed various challenges in recovering the microbial interactions. Firstly, the data is composed of discrete counts. Secondly, microbiome data entangles with the ``compositionality", which is a technicality imposed by the sequencing procedures. For instance, in 16S rRNA profiling, the ``sequencing depth", i.e. the total count of sequences in a sample, is pre-determined on the sequencing instrument, and is usually not on the same scale from sample to sample. This implies that the counts for each OTU in a sample carry only the information about their relative abundances instead of their absolute abundances. Thirdly, the microbiome data possesses ``high-dimensionality" in nature. With the resolution at the OTU level, it is likely that the number of OTU’s is far more than the number of samples in a biological experiment.

After one obtains the abundance data for the microbial species, marginal correlation analysis could be used to infer the interactions among microbes \citep{faust2012microbial}. Over the years, several methods have been developed to address the compositionality issue in the construction of correlation networks for microbiome data, such as SparCC \citep{friedman2012inferring} , CCLasso \citep{fang2015cclasso}, and REBECCA \citep{ban2015investigating}. All of those methods aimed to construct a covariance (correlation) matrix or network of the unknown absolute abundances (the positive, unconstrained true abundance of taxa). The high-dimensionality in microbiome data is tackled by imposing a sparsity constraint in the above three methods, with a correlation threshold for SparCC and an $L_1$-norm penalty for CCLasso and REBECCA.

All the above methods are built upon the marginal correlations between two microbial taxa, and they could lead to spurious correlations that are caused by confounding factors such as other taxa in the same community. Alternatively, interactions among taxa can be modeled through their conditional dependencies given the other taxa, which can eliminate the detection of spurious correlations. SPIEC-EASI was probably the first method that was proposed to estimate sparse microbial network based on conditional dependency \citep{kurtz2015sparse}. It first performs a central log-ratio (\textsc{clr}) transformation on the observed counts \citep{aitchison1986statistical}, and then apply graphical lasso \citep{yuan2007model, banerjee2008model, friedman2008sparse} to find the inverse covariance matrix of the transformed data. More recently, gCoda and CD-trace were developed to improve SPIEC-EASI by accounting for the compositionality property of microbiome data \citep{fang2017gcoda, yuan2019compositional}, both of which have been shown to possess better performance in terms of recovering the sparse microbial network than SPIEC-EASI.

It is worth noting that SPIEC-EASI, gCoda, and CD-trace are all built upon the \textsc{clr} transformation of the observed counts. Meanwhile, \cite{jiang2021microbial}  and \cite{tian2022compositional} proposed bias-corrected graphical lasso and compositional graphical lasso based on the additive log-ratio (\textsc{alr}) transformed data. In \textsc{alr} transformation, one needs to select a reference taxon and compute the log relative abundance of all other taxa with respect to the reference. One of the major concerns for the \textsc{alr} transformation is the robustness or invariance of the proposed method with respect to the choice of the reference taxon, which is not well studied in the literature.

In this paper, we first establish the reference-invariance property of estimating the sparse microbial network based on the \textsc{alr} transformed data. It shows that a submatrix of the inverse covariance matrix that correspond to the non-candidate-of-reference taxa is invariant with respect to the choice of the reference. Then, we propose a reference-invariant version of the compositional graphical lasso by modifying the penalty in its objective function, which only penalizes the invariant submatrix mentioned above. Additionally, we illustrate the reference-invariance property of the proposed method under a variety of simulation scenarios and also demonstrate its applicability and advantages by applying it to an oceanic microbiome data set.



\section{Methodology}
\label{Sec-3.2}

\subsection{Reference-Invariance Property}
\label{Sec-3.2.1}

Let $\bp = (p_1, \ldots, p_{K + 2})'$ denote a vector of compositional probabilities satisfying that $p_1 + \cdots + p_{K + 2} = 1$. The additive log-ratio (\textsc{alr}) transformation picks an entry of this vector as the reference and transforms the compositional vector using log ratios of each entry to the reference. Without loss of generality, suppose we pick the last entry as the reference, then the \textsc{alr} transformed vector becomes
\[ \bz = \left[ \log\left(\frac{p_1}{p_{K + 2}}\right), \ldots, \log \left(\frac{p_{K}}{p_{K + 2}} \right), \log \left(\frac{p_{K + 1}}{p_{K + 2}}\right) \right]'. \]
The transformed vector $\bz$ is often assumed to follow a multivariate continuous distribution with a mean vector $\bmu$ and a covariance matrix $\bSigma$. For example, $\bz \sim N(\bmu, \bSigma)$. Denote further the inverse covariance matrix $\bOmega = \bSigma^{-1}$.

Similarly, if we pick another entry as the reference, we can define another \textsc{alr}-transformed vector. For simplicity of illustration, suppose we choose the second last entry to be the reference and consider the following \textsc{alr} transformation
\[ \bz_p = \left[ \log\left(\frac{p_1}{p_{K + 1}}\right), \ldots, \log \left(\frac{p_{K}}{p_{K + 1}} \right), \log \left(\frac{p_{K + 2}}{p_{K + 1}}\right) \right]', \]
where the subscript $p$ denotes the ``permuted'' version of $\bz$. Similarly, define the mean vector of $\bz_p$ by $\bmu_p$, the covariance matrix by $\bSigma_p$, and the inverse covariance matrix by $\bOmega_p = \bSigma_p^{-1}$. 

A simple derivation implies that $\bz_p$ is a linear transformation of $\bz$ as $\bz_p = \bQ_p \bz$, where 
\begin{equation*}
\bQ_p = \left( \begin{matrix} \bI_{K} & \mathbf{-1} \\ \mathbf{0}' & -1 \end{matrix} \right)
\end{equation*}
with $\bI_{K}$ denoting the identity matrix, and $\mathbf{0}$ and $\mathbf{-1}$ denoting the column vectors with all $0$'s and all $-1$'s, respectively. It follows that $\bmu_p = \bQ_p \bmu$, $\bSigma_p = \bQ_p \bSigma \bQ_p'$, and $\bOmega_p = (\bQ_p')^{-1} \bOmega \bQ_p^{-1}$. It is also worth noting that $\bQ_p$ is an involutory matrix, i.e., $\bQ_p^{-1} = \bQ_p$.

The following theorem states the reference-invariance property of the inverse covariance matrix $\bOmega$ under the \textsc{alr} transformation.

\begin{thm} \label{Thm-3.1}
$\bOmega_{1:K, 1:K} =  \bOmega_{p, 1:K, 1:K}$, i.e. the $K \times K$ upper-left sub-matrix of the inverse covariance matrix of the \textsc{alr} transformed vector is invariant with respect to the choices of the $(K+2)$-th entry or the $(K+1)$-th entry as the reference.
\end{thm}

Theorem \ref{Thm-3.1} regards the reference-invariance property of the true value of the inverse covariance matrix $\bOmega$. It can also be extended to a class of estimators of $\bOmega$. Suppose we have i.i.d.\ observations of the compositional vectors $\bp_1, \ldots, \bp_n$, and consequently, their \textsc{alr} transformed counterparts $\bz_1, \ldots, \bz_n$. Then, we can construct an estimator of $\bSigma$, denoted by $\hat\bSigma$, based on the i.i.d.\ observations $\bz_1, \ldots, \bz_n$. Furthermore, we can construct an estimator of $\bOmega$, denoted by $\hat\bOmega$, by taking its inverse or generalized inverse. The following corollary presents the reference-invariance property for a class of such estimators.

\begin{cor} \label{Cor-3.1}
Suppose $\hat\bSigma_p = \bQ_p \hat\bSigma \bQ_p'$ and both $\hat\bSigma_p$ and $\hat\bSigma$ are invertible. Let $\hat\bOmega = \hat\bSigma^{-1}$ and $\hat\bOmega_p = \hat\bSigma_p^{-1}$ be their inverse matrices. Then, $\hat\bOmega_{1:K, 1:K} = \hat\bOmega_{p, 1:K, 1:K}$, i.e. the $K \times K$ upper-left sub-matrix of the estimated inverse covariance matrix of the \textsc{alr} transformed vector is invariant with respect to the choices of the $(K+2)$-entry or the $(K+1)$-th entry as the reference.
\end{cor}

The above results imply an important property for the additive log-ratio transformation in the compositional data analysis. It can be extended to a more general situation as follows. In general, suppose we have selected a set of entries $\bp_{\mathcal{R}}$ as ``candidate references'' in a compositional vector $\bp$ and write $\bp = (\bp_{\mathcal{R}^c}', \bp_{\mathcal{R}}')'$. Then, for any \textsc{alr} transformed vector $\bz$ based on a reference in the set of candidate references $\bp_{\mathcal{R}}$, the $|\mathcal{R}^c| \times |\mathcal{R}^c|$ upper-left sub-matrix of the (estimated) inverse covariance matrix of $\bz$ is invariant with respect to the choice of the reference. 

In the following subsections, we will incorporate the reference-invariance property into the estimation of a sparse inverse covariance matrix for compositional count data, such as the OTU abundance data in microbiome research.

\subsection{Logistic Normal Multinomial Model}
\label{Sec-3.2.2}

Consider an OTU abundance data set with $n$ independent samples, each of which composes observed counts of $K + 2$ taxa, denoted by $\bx_i = (x_{i,1}, \ldots, x_{i,K + 2})'$ for the $i$-th sample, $i = 1,\ldots,n$. Due to the compositional property of the data, the sum of all counts for each sample $i$ is a fixed number, denoted by $M_i$. Naturally, a multinomial distribution is imposed on the observed counts as
\begin{equation} 
\bx_i | \bp_i \sim \text{Multinomial}(M_i, \bp_i), \label{multinomial.comp.glasso}
\end{equation}
where $\bp_i = (p_{i,1}, \ldots, p_{i,K + 2})'$ are the multinomial probabilities with $\sum_{k=1}^{K + 2} p_{i,k} = 1$.

In addition, we choose one taxon, without loss of generality, the $(K + 2)$-th taxon as the reference and then apply the \textsc{alr} transformation \citep{aitchison1986statistical} on the multinomial probabilities as follows
\begin{equation}
\mathbf{z}_i =  \left[ \log \left(\frac{p_{i,1}}{p_{i, K + 2}}\right), \ldots, \log \left(\frac{p_{i, K}}{p_{i, K + 2}}\right), \log \left(\frac{p_{i, K + 1}}{p_{i, K + 2}}\right) \right]',\ i = 1,\ldots,n. \label{log.ratio.transformation.comp.glasso}
\end{equation}
Further assume that $\mathbf{z}_i$'s follow an i.i.d. multivariate normal distribution
\begin{equation}
\mathbf{z}_i \stackrel{iid}{\sim} N (\bmu, \bSigma),\ i=1,\ldots,n, \label{logistic.normal.comp.glasso}
\end{equation}
where $\bmu$ is the mean, $\bSigma$ is the covariance matrix, and $\bOmega = \bSigma^{-1}$ is the inverse covariance matrix. The above model in (\ref{multinomial.comp.glasso})--(\ref{logistic.normal.comp.glasso}) is called a logistic normal multinomial model and has been applied to analyze the microbiome abundance data \citep{xia2013logistic}. 

\cite{tian2022compositional} proposed a method called compositional graphical lasso that aims to find a sparse estimator of the inverse covariance matrix $\bOmega$, in which the following objective function is minimized
\begin{align}
\ell(\bz_1,\ldots,\bz_n,\bmu,\bOmega) ={}& -\frac1n\sum_{i=1}^n \left[\bx_{i, -(K+2)} ' \bz_i - M_i \log\{\mathbf{1}' \exp(\bz_i) + 1\}\right] \notag \\
& -\frac12 \log[\det(\bOmega)] + \frac{1}{2n} \sum_{i=1}^n (\bz_i - \bmu)' \bOmega (\bz_i - \bmu) + \lambda \|\bOmega\|_1, \label{objective.function.comp.glasso}
\end{align}
where $\bx_{i, -(K+2)} = (x_{i,1}, \ldots, x_{i, K+1})'$ and $\mathbf{1} = (1, \ldots, 1)'$. The above objective function has two parts: The first term in (\ref{objective.function.comp.glasso}) is the negative log-likelihood of the multinomial distribution in (\ref{multinomial.comp.glasso}) and the remaining terms are the regular objective function of graphical lasso for the multivariate normal distribution in (\ref{logistic.normal.comp.glasso}) regarding $\bz_1,\ldots, \bz_n$ as known quantities.

\subsection{Reference-Invariant Objective Function}
\label{Sec-3.2.3}

Similar to \ref{Sec-3.2.1}, if we choose another taxon, for simplicity of illustration, the $(K + 1)$-th taxon as the reference, then the \textsc{alr} transformation in (\ref{log.ratio.transformation.comp.glasso}) becomes
\[ \mathbf{z}_{i, p} =  \left[ \log \left(\frac{p_{i,1}}{p_{i, K + 1}}\right), \ldots, \log \left(\frac{p_{i, K}}{p_{i, K + 1}}\right), \log \left(\frac{p_{i, K + 2}}{p_{i, K + 1}}\right) \right]'.\]
As in Sections \ref{Sec-3.2.1}, $\mathbf{z}_{i, p} = \bQ_p \mathbf{z}_i$. Therefore, $\mathbf{z}_{i, p} \stackrel{iid}{\sim} N(\bmu_p, \bSigma_p)$, $i=1,\ldots,n$, where $\bmu_p = \bQ_p \bmu$, $\bSigma_p =  \bQ_p \bSigma \bQ_p'$, and $\bOmega_p = (\bQ_p')^{-1} \bOmega \bQ_p^{-1}$. The reference-invariance property in \ref{Sec-3.2.1} implies that $\bOmega_{1:K,1:K} =  \bOmega_{p, 1:K,1:K}$.

The different choice of the reference also leads to a different objective function for the compositional graphical lasso method (Comp-gLASSO) as follows
\begin{align}
\ell_p(\bz_{1, p},\ldots,\bz_{n, p},\bmu_p,\bOmega_p) = 
{}& -\frac1n\sum_{i=1}^n \left[\bx_{i, -(K+1)} ' \bz_{i,p} - M_i \log\{\mathbf{1}' \exp(\bz_{i,p}) + 1\}\right] \notag\\
& -\frac12 \log[\det(\bOmega_p)] + \frac{1}{2n} \sum_{i=1}^n (\bz_{i, p} - \bmu_p)' \bOmega_p(\bz_{i, p} - \bmu_p) + \lambda \|\bOmega_p\|_1, \label{objective.function.p}
\end{align}
where $\bx_{i, -(K+1)} = (x_{i,1}, \ldots, x_{i, K}, x_{i, K+2})'$ and $\mathbf{1} = (1, \ldots, 1)'$. Comparing (\ref{objective.function.comp.glasso}) and (\ref{objective.function.p}), their first terms are the same as they are both equal to the negative log-likelihood of the multinomial distribution: $-\frac1n \sum_{i=1}^n \sum_{k=1}^{K+2} x_{i,k} \log p_{i,k}$. In addition, from \citet{aitchison1986statistical}, $\det(\bOmega) = \det(\bOmega_p)$ and $\sum_{i=1}^n (\bz_i - \bmu)' \bOmega (\bz_i - \bmu) = \sum_{i=1}^n (\bz_{i, p} - \bmu_p)' \bOmega_p (\bz_{i, p} - \bmu_p)$ as known properties of the \textsc{alr} transformation. However, the $L_1$ penalties in (\ref{objective.function.comp.glasso}) and (\ref{objective.function.p}) are different because $\bOmega$ is not necessarily equal to $\bOmega_p$. The reference-invariance property only implies that $\bOmega_{1:K,1:K} = \bOmega_{p,1:K,1:K}$. 

Motivated by the reference-invariance property, we can impose the $L_1$ penalties only on the invariant entries of $\bOmega$ instead of all entries of $\bOmega$ as in (\ref{objective.function.comp.glasso}), which leads to 
\begin{align}
\ell_{inv}(\bz_1,\ldots,\bz_n,\bmu,\bOmega) ={}& -\frac1n\sum_{i=1}^n \left[\bx_{i, -(K+2)} ' \bz_i - M_i \log\{\mathbf{1}' \exp(\bz_i) + 1\}\right] \notag \\
& -\frac12 \log[\det(\bOmega)] + \frac{1}{2n} \sum_{i=1}^n (\bz_i - \bmu)' \bOmega (\bz_i - \bmu) + \lambda \|\bOmega_{1:K,1:K}\|_1. \label{invariant.objective.function}
\end{align}
With the previous arguments, we showed that $\ell_{\text{inv}}(\bz_1,\ldots,\bz_n,\bmu,\bOmega)$ is reference-invariant; in other words, the objective function $\ell_{\text{inv}}$ stays the same regardless of whether the $(K + 1)$-th or the $(K + 2)$-th taxa is selected as the reference. This is summarized in the following theorem.

\begin{thm} \label{Thm-3.2}
If $\bz_{i,p} = \bQ_p \bz_i$ for $i = 1,\ldots,n$, $\bmu_p = \bQ_p \bmu$, and $\bOmega_p = (\bQ_p')^{-1} \bOmega \bQ_p^{-1}$, then $\ell_{inv}(\bz_1,\ldots,\bz_n,\bmu,\bOmega) = \ell_{inv, p}(\bz_{1,p},\ldots,\bz_{n,p},\bmu_p,\bOmega_p)$.
\end{thm}

We call $\ell_{inv}(\bz_1,\ldots,\bz_n,\bmu,\bOmega)$ in (\ref{invariant.objective.function}) the reference-invariant compositional graphical lasso objective function. To obtain a sparse estimator of $\bOmega$, we minimize the objective function $\ell_{inv}(\bz_1,\ldots,\bz_n,\bmu,\bOmega)$ with respect to $\bz_1,\ldots,\bz_n$, $\bmu$, and $\bOmega$. We call this estimation approach the reference-invariant compositional graphical lasso (Inv-Comp-gLASSO) method.

In general, suppose we have selected a set of taxa $\bx_{\mathcal{R}}$ as ``candidate references'' and write $\bx = (\bx_{\mathcal{R}^c}', \bx_{\mathcal{R}}')'$. Then, the reference-invariant objective function becomes
\begin{align}
\ell_{\text{inv}}(\bz_1,\ldots,\bz_n,\bmu,\bOmega) ={}& -\frac1n\sum_{i=1}^n \left[\bx_{i, \mathcal{R}^c} ' \bz_i - M_i \log\{\mathbf{1}' \exp(\bz_i) + 1\}\right] \notag \\
& -\frac12 \log[\det(\bOmega)] + \frac{1}{2n} \sum_{i=1}^n (\bz_i - \bmu)' \bOmega (\bz_i - \bmu) + \lambda \|\bOmega_{\mathcal{R}^c, \mathcal{R}^c}\|_1. \label{invariant.objective.function.2}
\end{align}
In other words, (\ref{invariant.objective.function.2}) is invariant regardless of which reference is selected in the set of candidate references, and so is the invariant part of its minimizer.

It is noteworthy that the trick we played in defining the reference-invariant version of Comp-gLASSO is to revise the penalty term from the regular lasso penalty on the whole inverse covariance matrix to that only on the invariant part of the inverse covariance matrix. Using the same trick, we can define the reference-invariant version of other methods such as reference-invariant graphical lasso (Inv-gLASSO). The objective function of Inv-gLASSO is defined as follows when $\bz_1, \ldots, \bz_n$ are observed instead of $\bx_1, \ldots, \bx_n$:
\begin{equation}
\ell_{\text{inv}}(\bmu,\bOmega) = -\frac12 \log[\det(\bOmega)] + \frac{1}{2n} \sum_{i=1}^n (\bz_i - \bmu)' \bOmega (\bz_i - \bmu) + \lambda \|\bOmega_{\mathcal{R}^c, \mathcal{R}^c}\|_1. \label{Inv-gLASSO}
\end{equation}

The objective function (\ref{invariant.objective.function.2}) includes naturally three sets of parameters $(\bz_1,\ldots,\bz_n)$, $\bmu$, and $\bOmega$, which motivates us to apply a block coordinate descent algorithm. A block coordinate descent algorithm minimizes the objective function iteratively for each set of parameters given the other sets. Given the initial values $(\bz_1^{(0)},\ldots,\bz_n^{(0)})$, $\bmu^{(0)}$, and $\bOmega^{(0)}$, a block coordinate algorithm repeats the following steps cyclically for iteration $t= 0,1,2,\ldots$ until the algorithm converges.
\begin{enumerate}
\item Given $\bmu^{(t)}$ and $\bOmega^{(t)}$, find $(\bz_1^{(t+1)},\ldots,\bz_n^{(t+1)})$ that maximizes (\ref{invariant.objective.function.2}).
\item Given $(\bz_1^{(t+1)},\ldots,\bz_n^{(t+1)})$ and $\bOmega^{(t)}$, find $\bmu^{(t+1)}$ that maximizes (\ref{invariant.objective.function.2}).
\item Given $(\bz_1^{(t+1)},\ldots,\bz_n^{(t+1)})$ and $\bmu^{(t+1)}$, find $\bOmega^{(t+1)}$ that maximizes (\ref{invariant.objective.function.2}).
\end{enumerate}

Except for the mild modification on the objective function, the details of the algorithm is essentially the same as the one described in \cite{tian2022compositional}.

\section{Simulation Study}
\label{Sec-3.3}

\subsection{Settings}
\label{Sec-3.3.1}

To illustrate the reference-invariance property under the aforementioned framework, we conduct a simulation study and evaluated the performance of Inv-Comp-gLASSO as well as Inv-gLASSO.

Following \cite{tian2022compositional}, we generate three types of inverse covariance matrices $\bOmega = (\omega_{kl})_{1\le k,l \le K + 1}$ as follows:
\begin{enumerate}
\item \textbf{Chain}: $\omega_{kk} = 1.5$, $\omega_{kl} = 0.5$ if $|k - l| = 1$, and $\Omega_{kl} = 0$ if $|k - l| > 1$. Every node is connected to the adjacent node(s), and therefore the degree is $2$ for all but two nodes. 
\item \textbf{Random}: $\omega_{kl} = 1$ with probability $3/ (K + 1)$ for $k \ne l$. Every two nodes are connected with a fixed probability, and the expected degree is the same for all nodes. 
\item \textbf{Hub}: Nodes are randomly partitioned into $\lceil (K + 1) / 20 \rceil$ groups, and there's one ``hub node" in each group. For the other nodes in the group, they are only connected to the hub node but not each other. There's no connection among groups. The degree of connectedness is much higher for the hub nodes 
, and is $1$ for the rest of the nodes.
\end{enumerate}

In the simulations, we also vary two other factors that play a crucial role in the performances of the methods:
\begin{enumerate}
\item \textbf{Sequencing depth}. $M_i$'s are simulated from $\text{Uniform}(20K, 40K)$ or $\text{Uniform}(100K, 200K)$, denoted by ``low" and ``high" sequencing depth.
\item \textbf{Compositional variation}. For each aforementioned inverse covariance matrix $\bOmega$ (``low" compositional variation), we also divide each of them by a factor of $5$ to obtain another set of inverse covariance matrices, i.e., $\bOmega$/5 (``high" compositional variation). 

\end{enumerate}

The data are simulated from the logistic normal multinomial distribution in (\ref{multinomial.comp.glasso})--(\ref{logistic.normal.comp.glasso}). In detail, $\bz_i \sim N (\bmu, \bSigma)$ are first generated independently for $i = 1,\ldots,n$; then, the softmax transformation (the inverse of the \textsc{alr} transformation) was applied to get the multinomial probabilities $\bp_i$ with the $(K + 2)$-th entry serving as the true reference; last, the multinomial random variables $\bx_i$ were simulated from $Multinomial(M_i; \bp_i)$, for $i = 1,\ldots,n$. We set $n = 100$ and $K = 49$ throughout the simulations. 

The simulation results are based on 100 replicates of the simulated data. Both Inv-Comp-gLASSO and Inv-gLASSO are applied with two choices of reference, the $(K+1)$-th entry serving as the false reference and the $(K+2)$-th entry serving as the true reference. and only the reference-invariant sub-network $\bOmega_{1:K,1:K}$ is used in the evaluations. For Inv-gLASSO, we estimate $\bp_1,\ldots,\bp_n$ with $\bx_1/M_1, \ldots, \bx_n/M_n$, and performe the \textsc{alr} transformation to get the estimates of $\bz_1,\ldots,\bz_n$, which are denoted by $\tilde\bz_1,\ldots,\tilde\bz_n$. We then apply Inv-gLASSO to $\tilde\bz_1,\ldots,\tilde\bz_n$ directly to find the inverse covariance matrix $\bOmega$, which also serves as the starting value for Inv-Comp-gLASSO. For both methods, we implement them with a common sequence of $70$ tuning parameters of $\lambda$.

We empirically validate the invariance property of Inv-Comp-gLASSO and Inv-gLASSO by comparing the estimators of the two sub-networks $\bOmega_{1:K,1:K}$ resulted from choosing the true and false reference separately in each method, which have been shown to be theoretically invariant in Section \ref{Sec-3.2}. The comparison is assessed under four criteria as follows.

\begin{enumerate}
\item Normalized Manhattan Similarity

For two matrices $\mathbf{A}$ and $\mathbf{B}$, we define the normalized Manhattan similarity (NMS) as 
\[\text{NMS}(\mathbf{A},\mathbf{B}) = 1 - \frac{ \norm{\mathbf{A } - \mathbf{B}}_1}{\norm{\mathbf{A}}_1 + \norm{\mathbf{B}}_1},\]
where $\norm{\cdot}_1$ represents the entrywise $L_1$ norm of a matrix. Note that $0 \leq \text{NMS} \leq 1$ due to the non-negativity of norms and the triangle inequality.

\item Jaccard Index\\
For two networks with the same nodes, denote their set of edges by $\mathcal{A}$ and $\mathcal{B}$. Then the Jaccard Index \citep{jaccard1901distribution} is defined as follows:
\[J(\mathcal{A}, \mathcal{B}) = \frac{|\mathcal{A} \cap \mathcal{B}|}{|\mathcal{A} \cup \mathcal{B}|}. \]
Obviously, it also holds that $0 \leq J(\mathcal{A}, \mathcal{B}) \leq 1$.

\item Normalized Hamming Similarity\\
In the context of network comparison, the normalized Hamming similarity for two adjacency matrices $\mathbf{A}$ and $\mathbf{B}$ with the same $N$ nodes are defined as follows \citep{hamming1950error} 
\[H(\mathbf{A}, \mathbf{B}) = 1 - \frac{\norm{ \mathbf{A} - \mathbf{B}}_1} {N(N-1)},\]
where $\norm{\cdot}_1$ denotes the entrywise $L_1$ norm of a matrix. Since there are at most $N(N-1)$ $1$'s in an adjacency matrix with $N$ nodes, this metric is also between $0$ and $1$.

\item ROC Curve \\
The ROC curves on which true positive rate (TPR) and false positive rate (FPR) are plotted and also compared between the choices of true and false reference.


\end{enumerate}

\subsection{Results}
\label{Sec-3.3.2}

\subsubsection{Normalized Manhattan Similarity}
\begin{figure}[htbp]
\centering
\includegraphics[scale=0.425]{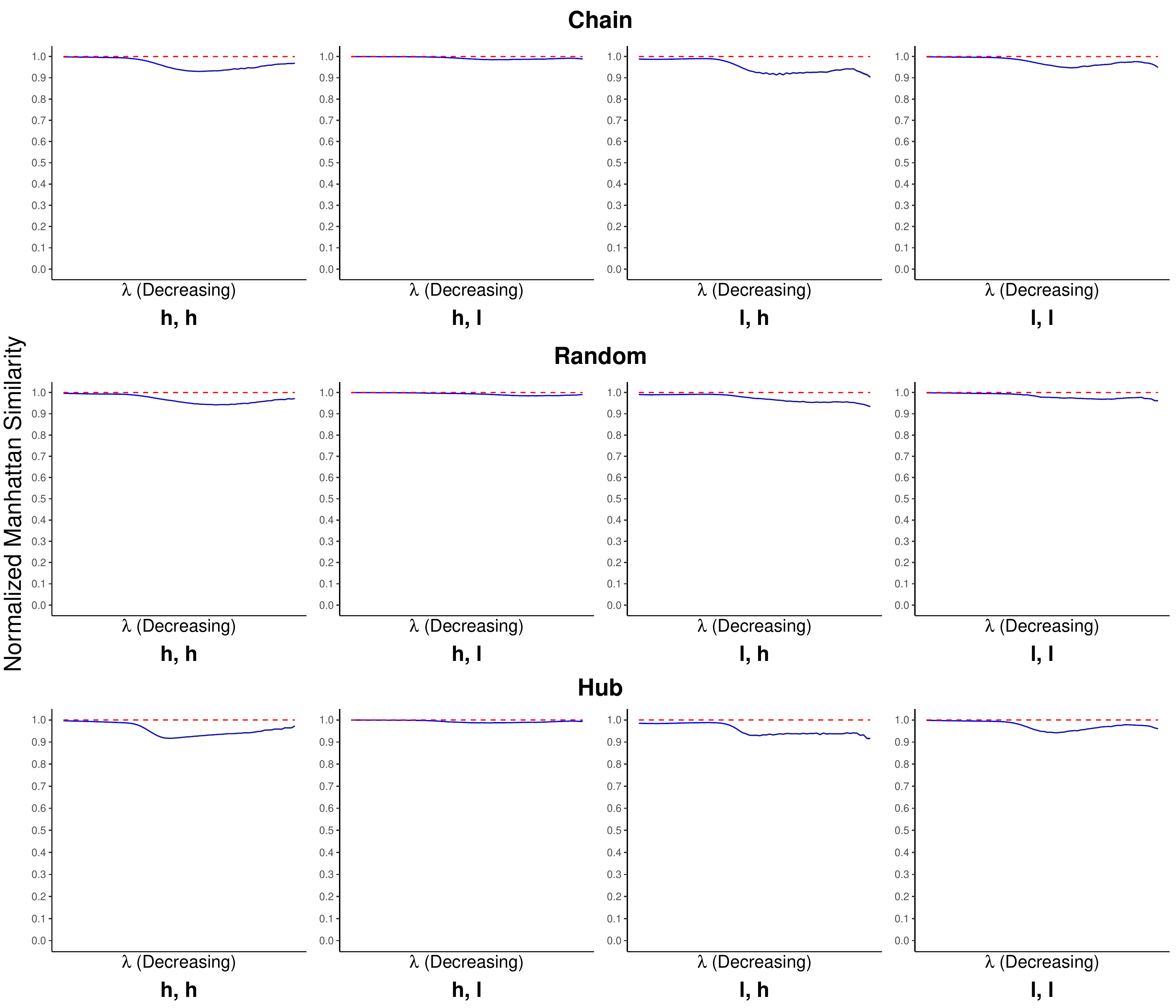}
\caption{Normalized Manhattan similarity between the two estimated inverse covariance matrices with true and false references. Solid blue: Inv-Comp-gLASSO; dashed red: Inv-gLASSO. $\mathbf{h, h}$: high sequencing depth, high compositional variation; $\mathbf{h, l}$: high sequencing depth, low compositional variation; $\mathbf{l, h}$: low sequencing depth, high compositional variation; $\mathbf{l, l}$: low sequencing depth, low compositional variation.}
\label{Normalized Manhattan Similarity}
\end{figure}

The normalized Manhattan similarity serves as a direct measure of the similarity between the two estimated inverse covariance matrices with different choices of the reference. On a sequence of $70$ tuning parameter $\lambda$'s, we calculated the normalized Manhattan similarity between the two estimated inverse covariance matrices with the true and false references separately. Figure \ref{Normalized Manhattan Similarity} shows the average normalized Manhattan similarity over 100 replicates along with standard error bars.
\label{Sec-3.3.2.1}

We can see from Figure \ref{Normalized Manhattan Similarity} that, the normalized Manhattan similarity for Inv-gLASSO stays close to 1 in all settings, throughout the whole sequence of tuning parameters. On the other hand, there are some fluctuations in the same metric from Inv-Comp-gLASSO, although most values stay higher than 0.9. Empirically, the two estimated matrices are numerically identical for Inv-gLASSO and close for Inv-Comp-gLASSO. A potential reason why the invariance of Inv-gLASSO is numerically more evident than that of Inv-Comp-gLASSO is that Inv-gLASSO is a convex optimization while Inv-Comp-gLASSO is not necessarily convex \citep{tian2022compositional}. With different starting points (as we choose different references), Inv-Comp-gLASSO might result in different solutions as the algorithm is only guaranteed to converge to a stationary point. We refer to \cite{tian2022compositional} for more detailed discussion about the convexity and the convergence of the algorithm.

In addition, it is consistently observed that the normalized Manhattan similarity for Inv-Comp-gLASSO starts close to 1 when $\lambda$ is very large and gradually decreases with some fluctuations as $\lambda$ decreases. This is because the Inv-Comp-gLASSO objective function in (\ref{invariant.objective.function}) is solved by an iterative algorithm between graphical lasso to estimate $(\bmu, \bOmega)$ and Newton-Raphson to estimate $\bz_1,\ldots,\bz_n$, which can lead to more numerical errors depending on the number of iterations. Furthermore, the algorithm is implemented with warm start for a sequence of decreasing $\lambda$'s, i.e., the solution for the previous $\lambda$ value is used as the starting point for the current $\lambda$ value. With the accumulation of numerical errors, the numerical difference between the two estimated matrices becomes larger.

Among the simulation settings, we find the invariance property for Inv-Comp-gLASSO is most evidently supported by the numerical results in the ``high sequencing depth, low compositional variation'' setting, regardless of the network types. The normalized Manhattan similarity is very close to 1 for Inv-Comp-gLASSO throughout the sequence of tuning parameters. This is because the compositional probabilities $\bp_i$'s and thus the $\bz_i$'s are estimated accurately with high sequencing depth and low compositional variation in the first iteration of the Inv-Comp-gLASSO algorithm, which implies fewer iterations for the algorithm to converge and less numeric error accumulated during this process. On the other hand, the normalized Manhattan similarity is the lowest in the ``low sequencing depth, high compositional variation'' setting. It is due to a similar reason that it takes more iterations for the Inv-Comp-gLASSO algorithm to converge, accumulating more numerical errors. However, it is noteworthy that this is exactly the setting in which Inv-Comp-gLASSO has the most advantage over Inv-gLASSO in recovering the true network (see Section \ref{Sec-3.3.2.3} for their ROC curves).

\subsubsection{Jaccard Index and Normalized Hamming Similarity}
\label{Sec-3.3.2.2}

\begin{figure}[htbp]
\centering
\includegraphics[scale=0.425]{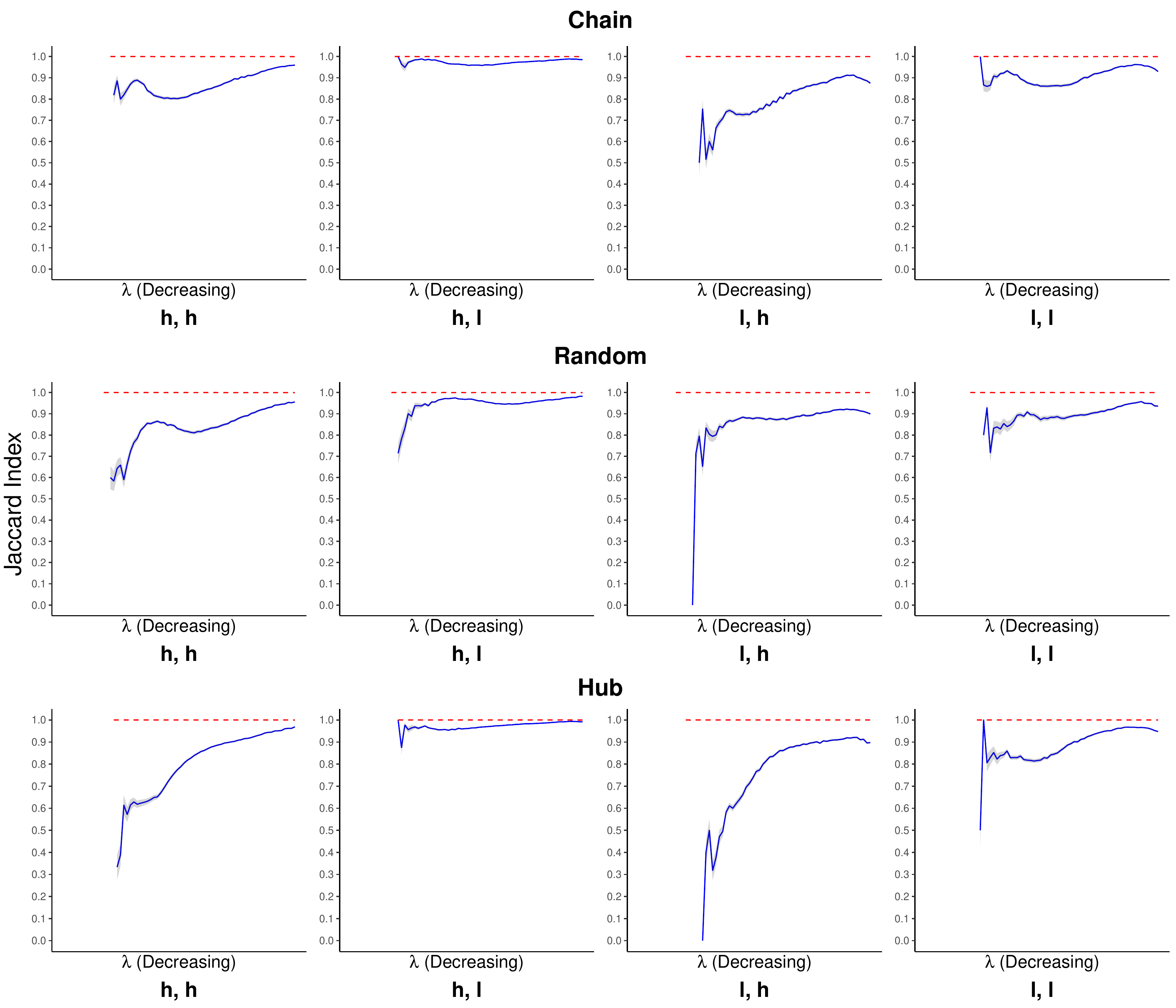}
\caption{Jaccard index between the two networks with true and false references. Solid blue: Inv-Comp-gLASSO; dashed red: Inv-gLASSO. $\mathbf{h, h}$: high sequencing depth, high compositional variation; $\mathbf{h, l}$: high sequencing depth, low compositional variation; $\mathbf{l, h}$: low sequencing depth, high compositional variation; $\mathbf{l, l}$: low sequencing depth, low compositional variation.}
\label{Jaccard Index}
\end{figure}

\begin{figure}[htbp]
\centering
\includegraphics[scale=0.425]{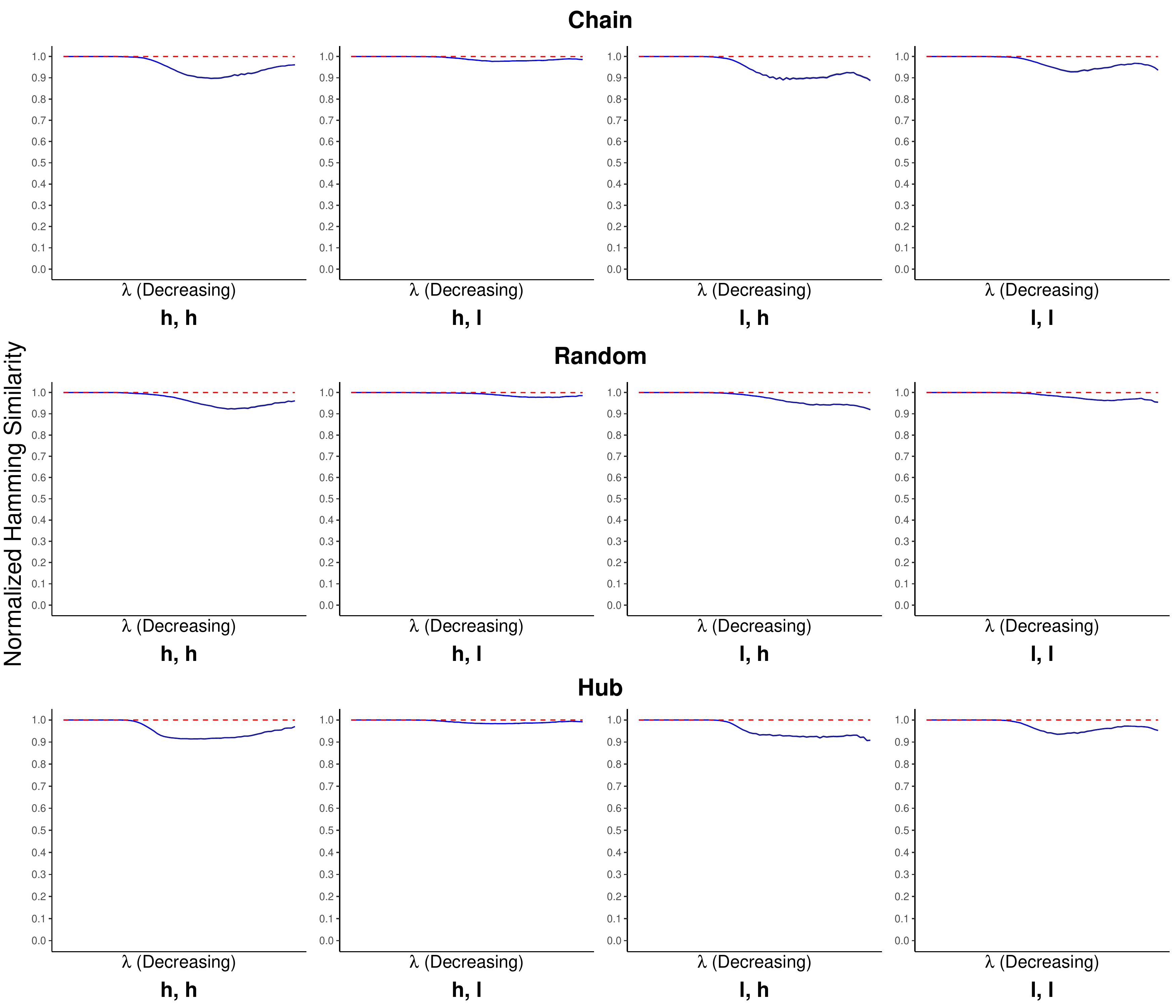}
\caption{Normalized Hamming similarity between the two networks with true and false references. Solid blue: Inv-Comp-gLASSO; dashed red: Inv-gLASSO. $\mathbf{h, h}$: high sequencing depth, high compositional variation; $\mathbf{h, l}$: high sequencing depth, low compositional variation; $\mathbf{l, h}$: low sequencing depth, high compositional variation; $\mathbf{l, l}$: low sequencing depth, low compositional variation.}
\label{Normalized Hamming Similarity}
\end{figure}

Compared to normalized Manhattan similarity that measures directly the similarity between two inverse covariane matrices, both Jaccard index and normalized Hamming similarity measure the similarity between two networks represented by the matrices because they only compared the adjacency matrices or the edges of the two networks. Again, on a sequence of $70$ tuning parameter $\lambda$'s, we computed the Jaccard index and the normalized Hamming similarity between the two networks with true and false references separately. Figures \ref{Jaccard Index} and \ref{Normalized Hamming Similarity} plot the average Jaccard index and the normalized Hamming similarity over 100 replicates along with standard error bars.

The results of normalized Hamming similarity in Figure \ref{Normalized Hamming Similarity} have a fairly similar pattern to those of normalized Manhattan similarity in Figure \ref{Normalized Manhattan Similarity} and thus can be similarly interpreted. We only focus on the results of Jaccard index in Figure \ref{Jaccard Index} here. The Jaccard index stays close to 1 for Inv-gLASSO, implying the identity between the two networks. Although the results of the Jaccard index for Inv-Comp-gLASSO look quite different from the other measures at the first glance, it actually implies a similar conclusion. First, we notice that there is no Jaccard index for the first few tuning parameters that are large enough. This is because the resultant network is empty with either true or false reference. Although two empty networks agree with each other perfectly, the Jaccard index is not well defined. Then, as Inv-Comp-gLASSO starts to pick up edges when $\lambda$ decreases, the Jaccard index is quite low in some settings, suggesting that the two networks are dissimilar. However, this is due to the fact the Jaccard index is a much more ``strict" similarity measure than the Hamming similarity. For example, for two networks with 100 possible total edges, if both networks only have one but a different edge, then the Jaccard index is $0$ while the normalized Hamming similarity is $0.98$. Finally, as the networks become denser, the Jaccard index increases quickly and stabilizes at a quite high value in most settings.

It is also notable that both the Jaccard index and the normalized Hamming similarity are relatively high in the ``high sequencing depth, low compositional variation'' setting and relatively low in the ``low sequencing depth, high compositional variation'' setting, which is consistent with the finding for the normalized Manhattan similarity.

\subsubsection{ROC Curves}
\label{Sec-3.3.2.3}
An ROC curve is plotted from the average of true positive rates and the average of false positive rates over 100 replicates. An ROC curve can be regarded as an indirect measure of the invariance (two networks possessing similar ROC curves is a necessary but not sufficient condition for the two networks to be similar). However, it is crucial to evaluate the algorithms with this criterion, since it answers the question: ``Does the performance of the algorithm depends on the choice of reference?"

\begin{figure}[h!]
\centering
\includegraphics[scale=0.425]{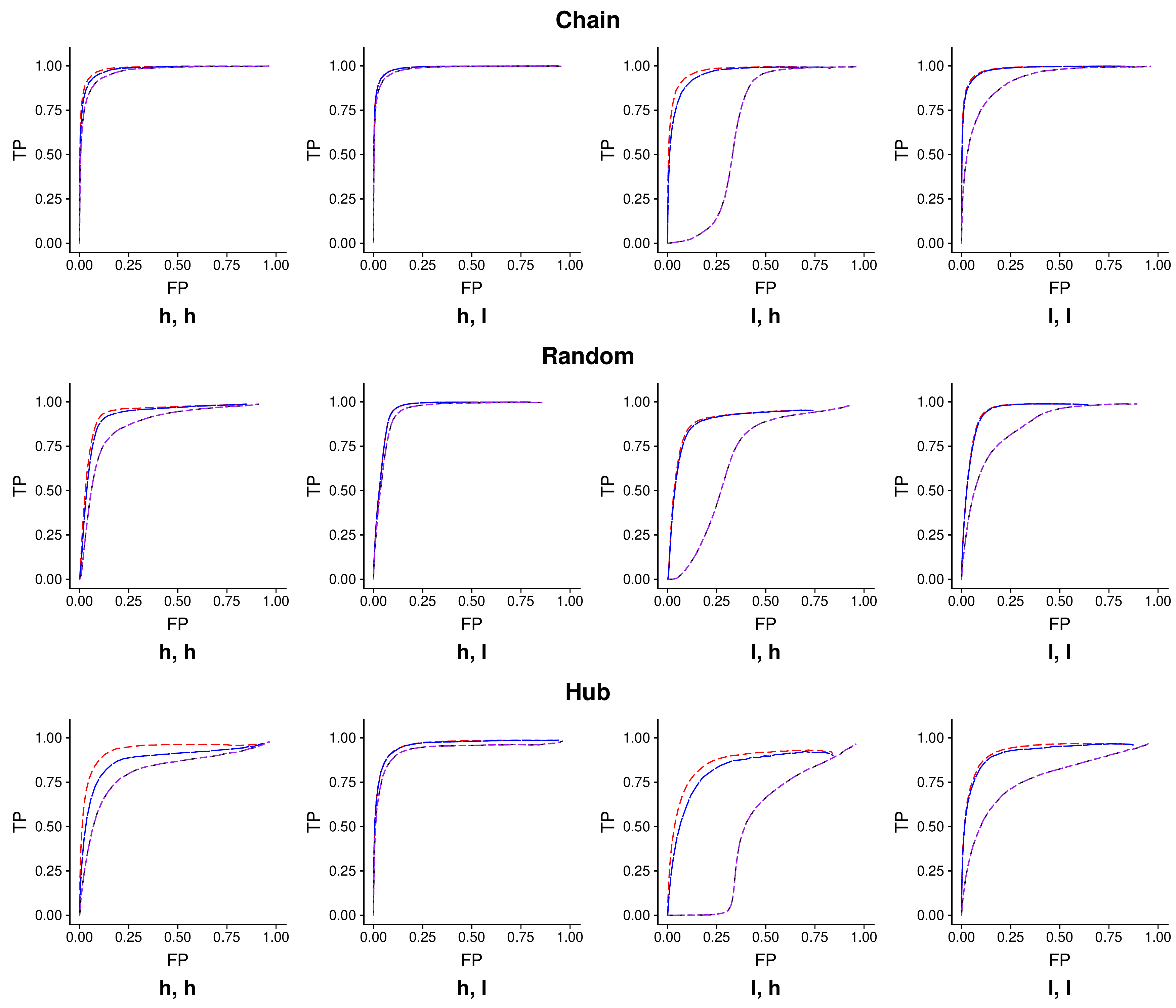}
\caption{ROC curves for Inv-Comp-gLASSO and Inv-gLASSO with true and false references. Long-dashed blue: Inv-Comp-gLASSO with true reference; dashed red: Inv-Comp-gLASSO with false reference; dashed-dotted purple: Inv-Comp-gLASSO with true reference; dotted black: Inv-Comp-gLASSO with false reference. $\mathbf{h, h}$: high sequencing depth and high compositional variation; $\mathbf{h, l}$: high sequencing depth and low compositional variation; $\mathbf{l, h}$: low sequencing depth and high compositional variation; $\mathbf{l, l}$: low sequencing depth and low compositional variation.}
\label{ROC_invar}
\end{figure}

We could see that the ROC curves from Inv-Comp-gLASSO, regardless of the choice of the reference, dominate the ones from Inv-gLASSO in all settings. The two ROC curves from Inv-gLASSO lay perfectly on top of each other, while the curves from Inv-Comp-gLASSO are also fairly close to each other. These empirical results validate the theoretical reference-invariance property for both methods. In addition, Inv-Comp-gLASSO has the most obvious advantage over Inv-gLASSO in the ``low sequencing depth, high compositional variation'' setting and they perform almost identically in the ``low sequencing depth, high compositional variation'' setting. Although the similarity measures are lower in the ``most favorable" setting for Inv-Comp-gLASSO (see Sections \ref{Sec-3.3.2.1} and \ref{Sec-3.3.2.2}), the ROC curves of the two networks from the method do not deviate too much from each other in this setting.

\section{Real Data}
\label{Sec-3.4}

To further validate the theoretical reference-invariance properties of Inv-Comp-gLASSO and Inv-gLASSO, we applied them to a dataset from the TARA Ocean project, in which the Tara Oceans consortium sampled both plankton and environmental data in 210 sites from the world oceans. The data collected was later analyzed using sequencing and imaging techniques. We downloaded the taxonomic data and the literature interactions from the TARA Ocean Project data repository (\url{https://doi.pangaea.de/10.1594/PANGAEA.843018}). As part of the TARA Oceans project, \cite{lima2015determinants} investigated the impact of both biotic and abiotic interactions in oceanic ecosystem. In this article, a literature-curated list of genus-level marine eukaryotic plankton interactions was generated by a panel of experts. 

Similar to \cite{tian2022compositional}, we focused the analysis on genus level and only kept the 81 genus involved in the literature-reported interactions. For computational simplicity, we removed the samples with too small reads ($< 100$). As a result, it leaves with 324 samples in the final preprocessed data. From the genera that were not reported in the literature, we chose two of them, Acrosphaera and Collosphaera, with the largest average relative abundances as the references. We then applied both Inv-Comp-gLASSO and Inv-gLASSO to the \textsc{alr}-transformed data with those two references, with a common sequence of tuning parameters. For each combination of a method and a reference, we also selected a tuning parameter that corresponds to the ``asymptotically sparsistent'' (the sparsest estimated network in the path that contains the true network asymptotically) network by StARS \citep{liu2010stability}.

\begin{figure}[htbp]
\centering
\includegraphics[scale=0.9]{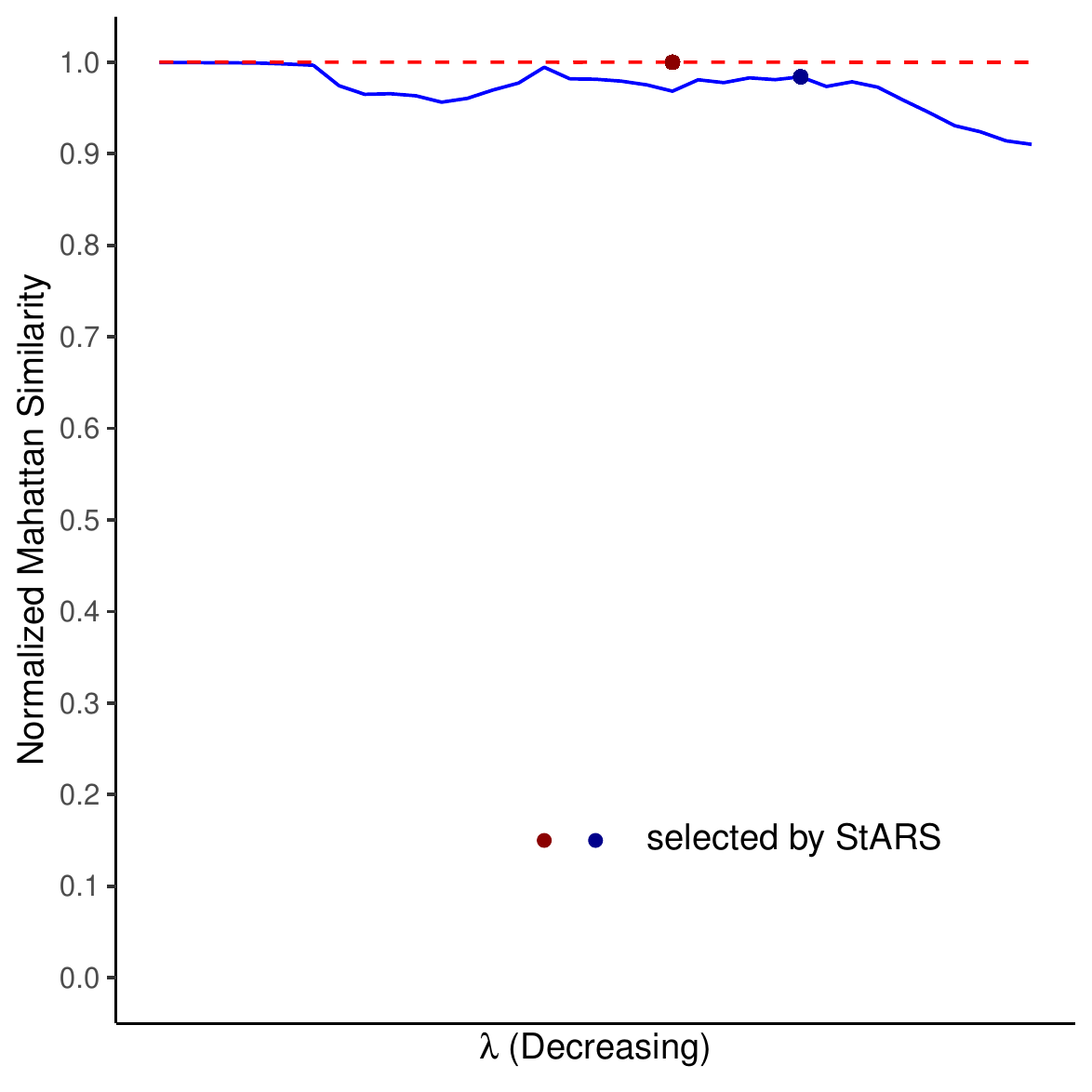}
\caption{Normalized Manhattan similarity between the two estimated inverse covariance matrices with the the two choices of reference, Acrosphaera and Collosphaera, from the TARA data. Solid blue: Inv-Comp-gLASSO; dashed red: Inv-gLASSO.}
\label{TARA_NMS_invar}
\end{figure}

\begin{figure}[htbp]
\centering
\includegraphics[scale=0.65]{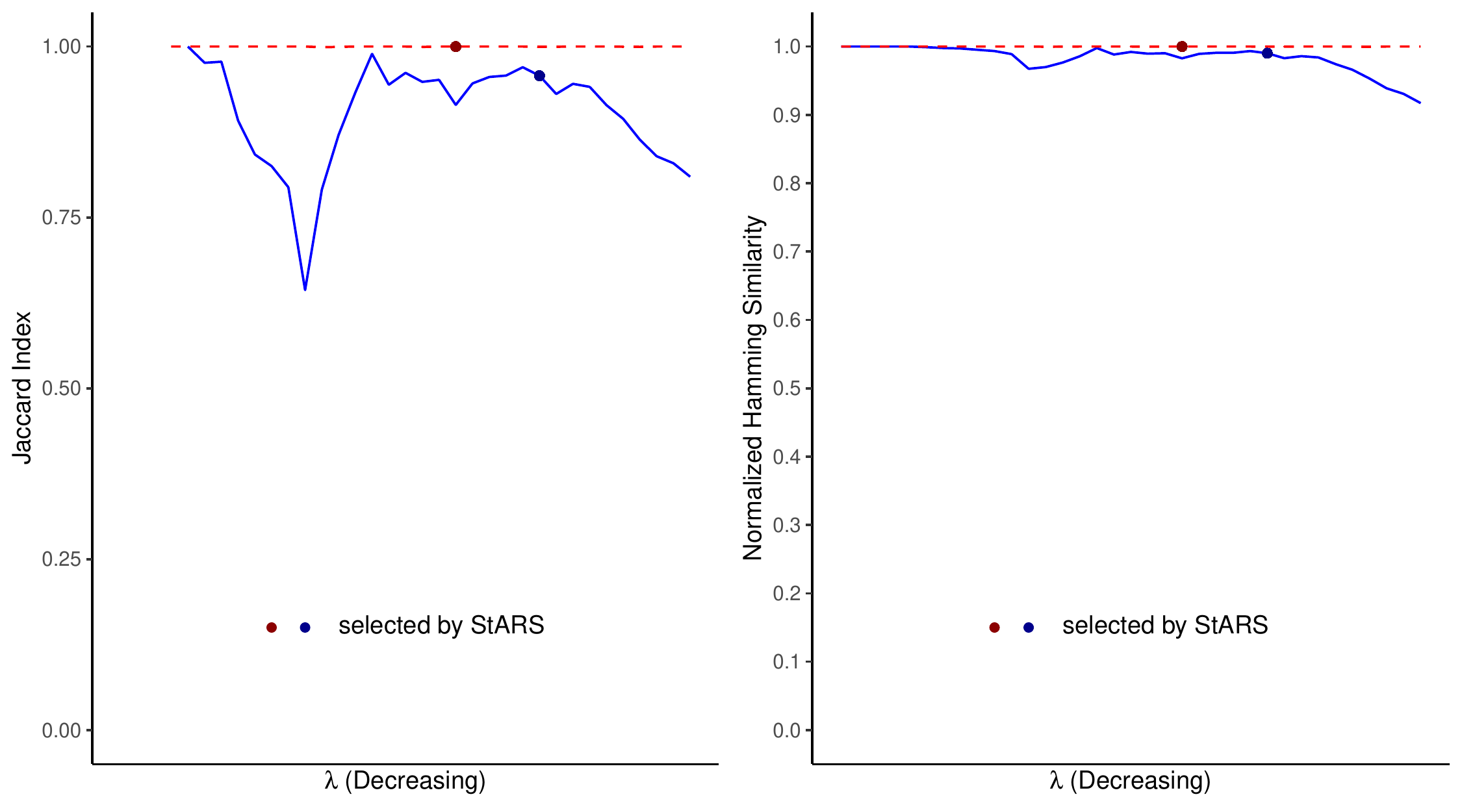}
\caption{Jaccard index and normalized Hamming similarity between the two estimated networks with the the two choices of reference, Acrosphaera and Collosphaera, from the TARA data. Solid blue: Inv-Comp-gLASSO; dashed red: Inv-gLASSO.}
\label{TARA_pair_invar}
\end{figure}

Figures \ref{TARA_NMS_invar} presents the normalized Manhattan similarity between two estimated inverse covariance matrices with the two choices of reference. Figures \ref{TARA_pair_invar} plotted the Jaccard index and the normalized Hamming similarity between the two networks represented by the two estimated inverse covariance matrices. We can see from Figure \ref{TARA_NMS_invar} and Figure \ref{TARA_pair_invar} that, all three similarity metrics stay steadily around $1$ for Inv-gLASSO throughout the sequence of $\lambda$'s. This agrees with our observation in simulations where Inv-gLASSO produced numerically identical inverse covariance matrices.

For Inv-Comp-gLASSO, the similarity scores start around $1$ (for normalized Manhattan similarity and normalized Hamming similarity) or non-existent (for Jaccard index) when the estimated networks are empty. Then, as $\lambda$ decreases, the estimated networks become denser, and the measures start to fluctuate and decline slightly at the end. In spite of the fluctuations, both normalized Manhattan similarity and normalized Hamming similarity stay above 0.9, while the lowest Jaccard index is about 0.64. As discussed earlier, Jaccard index is a stricter measure than normalized Hamming similarity. 

Within each method, StARS picked the same tuning parameter $\lambda$ regardless of the choice of the reference, as denoted by the red dot for Inv-gLASSO and the blue dot for Inv-Comp-gLASSO in Figures \ref{TARA_NMS_invar} and \ref{TARA_pair_invar}. In other words, the red and blue dots represent the tuning parameters corresponding to the final estimated networks selected by StARS. Again, the three similarity measures for the two final inverse covariance matrices or networks from Inv-gLASSO is almost 1, while the normalized Manhattan similarity, Jaccard index, and normalized Hamming similarity are $0.98$, $0.96$ and $0.99$ for Inv-Comp-gLASSO. All these high similarity scores imply the empirical invariance for Inv-gLASSO and Inv-Comp-gLASSO. Both methods result in invariant inverse covariance matrices and thus the corresponding networks with respect to the choices of the reference genus (Acrosphaera or Collosphaera) when applied to the TARA Ocean eukaryotic dataset.

\section{Discussion}
\label{Sec-3.5}
In this work, we established the reference-invariance property in sparse inverse covariance matrix estimation and network construction based on the \textsc{alr} transformed data. Then, we proposed the reference-invariant versions of the compositional graphical lasso and graphical lasso by modifying the penalty in their respective objective functions. In addition, we validated the reference-invariance property of the proposed methods empirically by applying them to various scenarios of simulations and a real TARA Ocean eukaryotic dataset.

It is noteworthy that the reference-invariance property is a general property for estimating the inverse covariance matrix based on the \textsc{alr} transformed data. We proposed reference-invariant versions of compositional graphical lasso and graphical lasso based on this property, however, one may revise other existing methods for inverse covariance matrix estimation based on the \textsc{alr} transformed data. The trick is to revise the objective function so that it becomes invariant with respect to the choice of the reference. Subsequently, the resultant inverse covariance matrix and network are expected to be reference-invariant both theoretically and empirically, the latter of which may depend on the algorithm that is used to optimize the reference-invariant objective function.

\bibliography{reference}
\bibliographystyle{asa}

\end{document}